\documentclass[sn-mathphys-num]{sn-jnl}

\usepackage{graphicx}%
\usepackage{multirow}%
\usepackage{amsmath,amssymb,amsfonts}%
\usepackage{slashed}
\usepackage{amsthm}%
\usepackage{mathrsfs}%
\usepackage[title]{appendix}%
\usepackage{xcolor}%
\usepackage{textcomp}%
\usepackage{manyfoot}%
\usepackage{booktabs}%
\usepackage{algorithm}%
\usepackage{algorithmicx}%
\usepackage{algpseudocode}%
\usepackage{listings}%
\usepackage{MnSymbol}
\theoremstyle{thmstyleone}%
\theoremstyle{thmstyletwo}%

\theoremstyle{thmstylethree}%

\begin{document}


\title[Article Title]{\bf{Exact Scalar and Fermionic Propagators in a Uniform Electric Background}}
\author*[1,2]{\fnm{{Emerson B. S.}} \sur{{Corrêa}}}\email{ecorreae@ufpa.br}

\affil*[1]{\orgdiv{Faculdade de Física}, \orgname{Universidade Federal do Pará - UFPA}, %
\postcode{66075-110},
\city{Belém}, \state{PA}, \country{Brazil}}

\affil*[2]{\orgdiv{Programa de Pós-Graduação em Física}, \orgname{Universidade Federal do Sul e Sudeste do Par\'a - Unifesspa},
\postcode{68505-080},
\city{Marabá}, \state{PA}, \country{Brazil}}

\abstract{

By implementing an analytical hybrid framework that combines the Ritus eigenfunction technique with Schwinger’s proper-time formalism, we construct exact propagators for charged scalar and fermionic fields in a constant, homogeneous, external electric field in four-dimensional Euclidean space. A central advantage of this approach is that the resulting propagators retain a four-dimensional momentum representation after the spectral summation, connecting the structure of the Ritus basis with the operator formulation of the proper-time method, without requiring the full traditional derivation of Schwinger’s approach and the coordinate restrictions originally associated with the Ritus method for fermions. The translationally non-invariant component emerging after the Laguerre-polynomial summation is isolated through a suitable change of variables, yielding translationally invariant kernels and maintaining the background field strength unchanged. As a consistency check, we successfully rederive Schwinger's seminal expression for the fermionic pair production rate.

}




\maketitle

\section{Introduction}

External electromagnetic fields provide interesting settings for investigating different aspects of quantum field theory (QFT), with applications ranging from vacuum polarization and effective actions~\cite{Schwinger,Dittrich,Dunne} to finite-temperature quantum electrodynamics (QED)~\cite{LoeweRojas1992,Elmfors1,Elmfors2}, lattice quantum chromodynamics (QCD)~\cite{Endrodi2025}, ultracold quantum gases~\cite{Experimento}, intense-laser physics~\cite{Fedotov2023,Li2023}, heavy-ion collisions~\cite{PRC-Ion}, and quantum systems in compactified geometries~\cite{PRDEmerson2023,EPJPlusEmerson2024}. For a pedagogical review of strong electromagnetic fields in QED and QCD, see~\cite{Hattori2023}. In this context, exact Green's functions provide essential building blocks for effective descriptions of charged quantum fields in prescribed external backgrounds.

In particular, magnetic fields have been extensively studied using exact propagators, with applications including dynamical symmetry breaking~\cite{Miransky} and modifications of the QCD phase diagram~\cite{DiagramaQCDMag}. In contrast, exact propagators in electric backgrounds have received comparatively less attention. They are particularly relevant for the study of nonperturbative phenomena, as considered in charged scalar systems at finite temperature~\cite{Tavares2024} and in studies of electric and inverse electric catalysis~\cite{Zamora}.
 
Recent investigations have revealed infrared sensitivities associated with the interplay between the zero-field and infinite-volume limits, as well as subtleties in defining thermodynamic equilibrium in the presence of electric fields~\cite{EndrodiMarko2022,EndrodiMarkoSandbote2026}. More recently, electric fields have been considered in formulations where nontrivial boundary wavefunctions are incorporated into self-energy-like contributions and the resummation associated with the Schwinger effect is encoded directly in the propagators~\cite{FukushimaMinato2026}.  External electric fields also have gained substantial interest in quantum information theory and relativistic quantum correlations~\cite{Wu2022_Electric, QIP2020, CTP2023}.

Exact propagators therefore play a central role in formulating such systems consistently and in constructing effective quantities. In this perspective, two methods for constructing propagators in external electromagnetic fields are widely used: Schwinger’s proper-time formalism~\cite{Schwinger} and Ritus’ eigenfunction method~\cite{Ritus,Ritus2}. Schwinger's approach expresses the propagator through the proper-time evolution of the corresponding differential operator, whereas the Ritus method constructs a field basis that diagonalizes the relevant operator. Both methods have been applied to charged Bose, Dirac, and vector fields, providing expressions for Green's functions in nontrivial backgrounds~\cite{Chodos,Lawrie,Khalilov,ElizaldePropRitus,Mexicanos,Igor,RBEFEmerson2015,CJPEmerson2022,RBEFEmerson2024,CMEmerson2022,Sanchez}. More recently, a hybrid construction combining the Ritus eigenfunction method with Schwinger's proper-time formalism was applied for spin-0 and spin-1/2 charged fields in a constant magnetic background~\cite{Correa2025}.

Methodologically, the field-mode decomposition under a uniform electric background shares instructive parallels with field quantization in curved spacetimes, such as Schwarzschild black hole geometries \cite{Wu2022_Curved, Wu2023_Entanglement}. In both settings, a strong external background induces a fundamental instability in the quantum vacuum, leading to particle creation via the the Schwinger mechanism or the Hawking effect, respectively. In curved geometries, information dynamics is evaluated by expanding fields in adapted modes across the event horizon. Analogously, in an electric background, the eigenfunctions $E_p(u)$ provide the exact basis that diagonalizes the wave operator. While black holes possess geometric event horizons that restrict physical accessibility to information, a constant electric field establishes an effective acceleration horizon for charged particles.

Despite these developments, the formulation combining the Ritus eigenfunction method and Schwinger’s proper-time formalism for charged Bose and Dirac fields in a constant electric background has, to the best of our knowledge, not yet been presented. This motivates the present work. We combine the Ritus and Schwinger methods to derive exact Green's functions for bosonic and fermionic charged fields in an external electric field. After carrying out the spectral sum, we obtain the propagators in a four-dimensional momentum representation. We also separate the coordinate-dependent Schwinger phase from the translationally invariant part of the Green's function, providing a representation suitable for subsequent applications to finite-temperature and compactified systems.

The primary objective of this work is to provide a new derivation in four-dimensional momentum space of the scalar and fermionic propagators in a uniform electric background, specifically through a hybrid Ritus-Schwinger scheme. We perform an exact spectral summation over the discrete oscillator levels (expressed via Laguerre and Hermite polynomials) to obtain closed-form expressions. Crucially, while the present derivation offers a distinct algebraic pathway, we demonstrate that our resulting Green's functions successfully recover the free propagators in the vanishing external field limit, as well as the established fermionic pair production rate expression, confirming both the rigor and accuracy of the hybrid framework.

The paper is organized as follows. In Sec.~II, we derive the exact Green's function for a charged spinless field by combining the Ritus basis with the proper-time representation. In Sec.~III, we extend the construction to the spin-1/2 field and obtain the corresponding propagator. In Sec. IV, we reobtain Schwinger's seminal result concerning the fermionic pair creation rate using the propagator computed with the hybrid technique.
 Conclusions are presented in Sec.~V. In Appendix A, we evaluated some traces useful in the context of the Schwinger mechanism. We use natural units, $c=\hbar=1$, and work throughout the paper in a four-dimensional Euclidean space, unless specified otherwise.
\section{Propagation in a Uniform Electric Background: Bose Field}
\subsection{Ritus representation of the scalar field}

We consider a charged scalar field in four-dimensional Euclidean space interacting with a uniform external electric background $\mathbf{{E}}$ oriented along the $z$-direction. The dynamics of the spinless field modes are governed by the covariant Klein--Gordon equation
\begin{eqnarray*}
\left(-D_{\mu}^{2} + m_{0}^{2}\right)\Phi\left(u\right) = 0,
\label{Eq.K-GcomCampo}
\end{eqnarray*}
with $u\equiv u^{\rho} = (\tau,x,y,z)$ representing the Euclidean four-coordinate, and the covariant derivative is defined via minimal coupling as $D_{\mu} = \partial_{\mu} - i e A_{\mu}^{ext}$, with $e$ denoting the electric charge of the scalar field.

To describe this uniform setup, we adopt the following gauge configuration in Euclidean space: $A_{\mu}^{ext} = (-z{E},0,0,0)$, where ${E}$ is a constant parameter. This potential directly yields the Euclidean external electric field ${\bf{E}} = -\,{\bf{\nabla}}{A}^{ext}_{0} = {E}\hat{z}$. The physical electric field $E_{\rm{M}}$ and time $t$ are defined in Minkowski spacetime. Consequently, the transition to the Euclidean domain is established through the identifications $t \, \rightarrow \,- \,i\,\tau$ and $E_{\rm{M}} \, \rightarrow \, + \, i \, {E}$, as discussed in Ref.~\cite{EndrodiMarkoSandbote2026}.

Within this framework, the scalar Green's function satisfies the inhomogeneous differential equation
\begin{eqnarray}
\left(D_{\mu}^{2} - m_{0}^{2}\right)G\left(u,u^{\prime},A\right) = -\,\delta^{4}\left(u-u^{\prime}\right).
\label{eqdeGreenBosons}
\end{eqnarray}

To determine this propagator, we first implement the eigenfunction technique established by Ritus in the 1970s. This approach proceeds by expanding the coordinate-space Green's function in terms of a complete basis set, $E_p(u)$, which diagonalizes the quadratic covariant derivative operator. The underlying eigenvalue problem is given by
\begin{eqnarray}
D_{\mu}^2 \,E_p = -\, p^2 \,E_p,
\label{eigenvalue}
\end{eqnarray}
where, under our prescribed gauge condition, the explicit form of the operator reads
\begin{eqnarray}
D_{\mu}^2  = \nabla^{2}_{4D} + 2\,i\,\Omega \,z\, \partial_{\tau} -\Omega^{2} z^{2},
\label{operator2}
\end{eqnarray}
with $\Omega \equiv e {E} >0$ defining the specific energy scale introduced by the constant external electric field. Once the complete set of eigenfunctions $E_p(u)$ is systematically derived, the full Green's function can be expressed via the decomposition
\begin{eqnarray}
G\left(u,u^{\prime},A\right) = \sumint dp \,E_{p}(u) \,\tilde{G}(p,A) \,E_{p}^{*}(u^{\prime}).
\label{Green}
\end{eqnarray}
By utilizing the standard completeness condition for the Ritus basis, namely,
\begin{eqnarray}
\sumint dp \,E_p (u) \, E^{*}_{p}(u^\prime) = \delta^4(u-u^\prime),
\label{fechamento}
\end{eqnarray}
and, upon substituting Eqs.~(\ref{eigenvalue}) and (\ref{Green}) into Eq.~(\ref{eqdeGreenBosons}) alongside the identity (\ref{fechamento}), the momentum-space scalar propagator is found to be
\begin{eqnarray}
\tilde{G}(p,A) = \frac{1}{p^2 + m_{0}^2}.
\label{propagmomenta}
\end{eqnarray}
The task of obtaining the coordinate-space propagator relies on evaluating the explicit forms of $E_p(u)$ along with their corresponding eigenvalues $p^2$. In this constant background context, Eq.~(\ref{Green}) coupled with Eq.~(\ref{propagmomenta}) delivers the appropriate representation for the scalar field propagator.

To decouple the differential equation along the direction of the field, we introduce a Fourier-type ansatz
\begin{eqnarray}
E_p = \mathcal{N} \exp\left[i\left(\Omega \, \zeta \, \tau + p_x x + p_y y \right)\right]Z(z).
\label{ansatze}
\end{eqnarray}
By applying this ansatz to the eigenvalue problem, i.e., Eq.~(\ref{eigenvalue}), along with the explicit form of the covariant operator in Eq.~(\ref{operator2}), the system reduces to a one-dimensional differential equation governing the spatial mode $Z(z)$:
\begin{eqnarray}
\left[\frac{d^{2}}{dz^{2}}-\Omega^2 (z+\zeta)^2 + (p^{2}-p_{x}^{2}-p_{y}^{2}) \right]Z(z) = 0.
\label{Hermite}
\end{eqnarray}

This relation displays the standard form of the Hermite differential equation. Physically acceptable, square-integrable solutions exist exclusively when the constant grouping is constrained to discrete values, enforcing the profile
$(p^{2}-p_{x}^{2}-p_{y}^{2}) \equiv \Omega \, (2\,\ell+1),$
where $\ell$ is a non-negative integer. This boundary requirement ensures that the functional behavior of $Z(z)$ remains regular and normalizable over the entire domain.

Consequently, the solutions along the $z$-coordinate direction can be neatly expressed in terms of normalized Hermite functions as
\begin{eqnarray}
{Z}_{\ell}(z) &=& (\Omega)^{1/4}\, h_{\ell}[\sqrt{\Omega}(z+\zeta)],           
\label{autofuncaoB}
\end{eqnarray}
where the functions $h_{\ell}[\sqrt{\Omega}(z+\zeta)]$ denote the normalized Hermite functions, defined by
\begin{eqnarray}
{h}_{\ell}({s}) &\equiv& \frac{1}{\sqrt{2^{\ell }\ell \,!\,\sqrt{\pi}}}\exp\left(-s^{2}/2\right) H_{\ell }\left(s\right).   
\label{FH}
\end{eqnarray}
The underlying algebraic properties of these polynomial forms satisfy the following standard orthonormality and completeness relations~\cite{alemao}:
\begin{eqnarray}
\left\lbrace
	\begin{array}{lcl}
   		\dfrac{}{}\int_{-\infty}^{\infty} \,{h}_{\ell}(s)\,{h}_{\ell^{\prime}}(s)\,ds, & = & \delta_{\ell,\ell^{\prime}}, \\
                                     \\
	\sum_{\ell=0}^{\infty} \dfrac{}{}	{h}_{\ell}(s)\,{h}_{\ell}(s^{\prime}) & = & \delta(s-s^{\prime}).
  	\end{array}
	\right.
    \label{chave}
\end{eqnarray}

By combining the ansatz introduced in Eq.~(\ref{ansatze}) with the orthonormality and completeness conditions presented in Eq.~(\ref{chave}), and invoking the standard Fourier integral representation of the Dirac delta function, the normalization condition required by Eq.~(\ref{fechamento}) is satisfied. This procedure establishes the amplitude factor as $|\mathcal{N}| = 1/\sqrt{(2\pi)^{3}}$, which uniquely determines the complete Ritus eigenfunctions:
\begin{eqnarray}
E_{p}(u)=\frac{\Omega^{1/4}}{\sqrt{(2\pi)^3}}\,\exp\left[i\left({\Omega \,\zeta\,\tau + p_{x}x + p_{y}y}\right)\right] \, h_{\ell}[\sqrt{\Omega}(z+\zeta)].
\label{RitusRepaginado}
\end{eqnarray}
 
With the complete set of eigenfunctions established, the coordinate-space scalar propagator can be expanded within the Ritus basis as
\begin{eqnarray}
G\left(u,u^{\prime},A\right) =\,\sum_{\ell=0}^{\infty}\, \int \, dp_x \,dp_y \,\Omega \, d\zeta   \, E_{p}(u) \,\tilde{G}(p,A)\,E_{p}^{*}(u^{\prime}),
\label{propRitus}
\end{eqnarray}
where $E_p(u)$ is given by Eq.~(\ref{RitusRepaginado}) and $\tilde{G}(p,A)$ is determined by Eq.~(\ref{propagmomenta}). The operational signature of the external electric background is carried by the discrete eigenvalue spectrum, which takes the form
\begin{eqnarray}
p^2 = p_x^2+p_y^2+\Omega(2\ell+1),\,\,\,\,\,\,\ell = 0,1,2,3,\cdots.
\label{pdois}
\end{eqnarray}
Here, the integer index $\ell$ labels the discrete eigenvalues of the covariant operator mapped within the $(\tau, z)$ coordinate sector. It is worth emphasizing that, unlike the magnetic scenario where these discrete states signify orbital Landau quantization, the spectrum in this electric configuration emerges from a kind of effective harmonic-oscillator potential.

After continuation back to real time $\tau \rightarrow  + i \, t$ and a real electric field, \( E\rightarrow - i\,E_{\rm{M}}\), the corresponding oscillator frequency becomes imaginary. The discrete spectrum obtained in the Euclidean formulation in Eq.~(\ref{pdois}) should not be interpreted as a discrete spectrum of the physical electric-field problem. In particular, for fixed \(p_x\) and \(p_y\), the corresponding Minkowski spectral structure is expected to become continuous.
\subsection{Schwinger representation of the scalar field}

We next recast the scalar propagator in the Schwinger proper-time formalism. Introducing the proper-time parameter $S$, we employ the operator identity
\begin{eqnarray}
\mathcal{O}^{-1} = \int_{0}^{\infty} \, dS \exp{\left(-S \, \mathcal{O}\right)},
\label{Operatorial}
\end{eqnarray}
and apply it to Eq.~(\ref{propRitus}), with
$\mathcal{O} \equiv \left[p_{x}^{2}+p_{y}^{2}+\Omega(2\ell+1)+m_{0}^{2}\right]$.
This yields the following proper-time representation of the coordinate-space propagator:
\begin{eqnarray}
G\left(u,u^{\prime},A\right) &=& 
\int_{0}^{\infty}  dS \,\int \frac{dp_x}{2\pi} \frac{dp_y}{2\pi} \frac{(\Omega)^{3/2}}{2\pi} \,\exp\left\{i\left[p_{x}\left(x-x^{\prime}\right)+p_y\left(y-y^{\prime}\right)\right]\right\} \nonumber \\
&\times&\,\exp\left[-S \left(p_{x}^{2}+p_{y}^{2}+m_{0}^{2}\right)\right] \mathcal{I},
\label{16}
\end{eqnarray}
where the isolated structure capturing the explicit $\ell$-summation and the auxiliary phase parameter $\zeta$ reads
\begin{eqnarray}
\mathcal{I} &\equiv& \sum_{\ell=0}^{\infty} \exp\left[-S\Omega(2\ell+1)\right] \int_{-\infty}^{\infty} d\zeta \exp\left[i\Omega\zeta(\tau-\tau')\right] h_{\ell}\left[\sqrt{\Omega}(z+\zeta)\right] h_{\ell}\left[\sqrt{\Omega}(z'+\zeta)\right]. \nonumber \\
\label{f}
\end{eqnarray}

\subsubsection*{Evaluation of the $\zeta$-Dependent Contribution}

To evaluate the quantity $\mathcal{I}$, we collect all terms in Eq.~(\ref{f}) that depend on the intermediate variable $\zeta$. These contributions originate from the phase factor and from the Gaussian factors contained in the Hermite functions defined in Eq.~(\ref{FH}). We therefore introduce the function $f(\zeta)$ to represent the complete exponential argument
\begin{equation}
f(\zeta) = -\frac{\Omega}{2}(z+\zeta)^2 - \frac{\Omega}{2}(z'+\zeta)^2 + i\Omega(\tau-\tau^{\prime})\zeta.
\end{equation}
Expanding the quadratic terms and grouping the expressions by powers of $\zeta$ yields:
\begin{equation}
f(\zeta) = -\Omega\zeta^2 - \Omega\left[ z + z' - i(\tau-\tau') \right]\zeta - \frac{\Omega}{2}\left(z^2 + z'^2\right).
\end{equation}
By factoring out $-\Omega$ from the terms containing $\zeta$, we can identify the linear coefficient to complete the square:
\begin{equation}
f(\zeta) = -\Omega \left\{ \zeta^2 + \left[ z + z' - i(\tau-\tau') \right]\zeta \right\} - \frac{\Omega}{2}\left(z^2 + z'^2\right).
\label{a}
\end{equation}
We now introduce the complex shift parameter $\mathcal{A}$, defined as:
\begin{equation}
\mathcal{A} \equiv \frac{(z+z') - i(\tau-\tau')}{2}.
\label{b}
\end{equation}
Using this definition, the quadratic expression in the braces can be rewritten as a perfect square plus a correction term:
\begin{equation}
\zeta^2 + 2\mathcal{A}\zeta = (\zeta + \mathcal{A})^2 - \mathcal{A}^2.
\label{c}
\end{equation}
Substituting Eqs.~(\ref{c}) and (\ref{b}), back into Eq.~(\ref{a}) splits the exponential argument into a shifted Gaussian in $\zeta$ and a residual term $R$:
\begin{equation}
f(\zeta) = -\Omega(\zeta + \mathcal{A})^2 + R,
\end{equation}
where the $\zeta$-independent term $R$ is given by:
\begin{equation}
R = \Omega\mathcal{A}^2 - \frac{\Omega}{2}\left(z^2 + z'^2\right).
\end{equation}
Expanding $\mathcal{A}^2$ explicitly under a common denominator, gives:
\begin{equation}
R = \frac{\Omega}{4} \left[ (z+z')^2 - 2i(z+z')(\tau-\tau') - (\tau-\tau')^2 - 2z^2 - 2z'^2 \right].
\end{equation}
Using the identity $(z+z')^2 - 2z^2 - 2z'^2 = -(z-z')^2$, the expression for $R$ simplifies directly to:
\begin{equation}
R = -\frac{\Omega}{4}\left[(z-z')^2 + (\tau-\tau')^2\right] - \frac{i\Omega}{2}(z+z')(\tau-\tau').
\end{equation}
Consequently, the full exponential argument within Eq.~(\ref{f}) maps identically onto the compact function $f(\zeta)$ reorganized by
\begin{equation}
f(\zeta) = -\Omega\left\{ \left[\zeta+\mathcal{A}\right]^2 +\frac{1}{4}\left[(z-z')^2+(\tau-\tau')^2\right] +\frac{i}{2}(z+z')(\tau-\tau') \right\}.
\end{equation}
Finally, the $\zeta$-independent factor $\exp(R)$ can be moved outside the integration sign, allowing the full integral $\mathrm{\mathcal{I}}$ to be written as:
\begin{align}
\mathrm{\mathcal{I}} = \sum_{\ell=0}^{\infty} \frac{\exp\left[-S\,\Omega \,(2\ell+1)\right]}{2^{\ell }\ell \,!\,\sqrt{\pi}} &\exp\left\{ -\frac{\Omega}{4}\left[(z-z')^2 + (\tau-\tau')^2\right] - \,i\frac{\Omega}{2}(z+z')(\tau-\tau') \right\} \nonumber \\
&\times \int_{-\infty}^{+\infty} d\zeta \exp\left[ -\Omega(\zeta + \mathcal{A})^2 \right] H_{\ell }\left[\sqrt{\Omega}(z+\zeta)\right] H_{\ell }\left[\sqrt{\Omega}(z^{\prime}+\zeta)\right].
\label{Xi44}
\end{align}

Making the variable change
\begin{eqnarray*}
\Theta=\sqrt{\Omega}\left(\zeta+\mathcal{A}\right),
\end{eqnarray*}
Eq.~(\ref{Xi44}) becomes
\begin{eqnarray}
\mathrm{\mathcal{I}}
&=&
\sum_{\ell=0}^{\infty}
\exp\left[{-S\Omega(2\ell+1)}\right]
\frac{1}{\sqrt{\Omega}\,2^\ell\ell!\sqrt{\pi}}
\exp\left[-\,
i\frac{\Omega}{2}(z+z')(\tau-\tau')
\right]
\exp\left\{
-\frac{\Omega}{4}
\left[(z-z')^2+(\tau-\tau')^2\right]
\right\}
\nonumber\\
&&\times
\int_{-\infty}^{+\infty}
d\Theta\,
\exp{(-\Theta^2)}
H_\ell\!\left[
\Theta+\sqrt{\Omega}(z-\mathcal{A})
\right]
H_\ell\!\left[
\Theta+\sqrt{\Omega}(z'-\mathcal{A})
\right].
\label{Xi3}
\end{eqnarray}

Using the well-known identity
\begin{eqnarray*}
\int_{-\infty}^{+\infty}
d\Theta\,
\exp{(-\Theta^2)}
H_\ell(\Theta+b)
H_{\ell'}(\Theta+b')
=
2^{\ell'}\sqrt{\pi}\,
\ell!\,
(b')^{\ell-\ell'}
L_\ell^{\,\ell'-\ell}
(-2\,b\,b'),
\end{eqnarray*}
we obtain
\begin{eqnarray}
\mathrm{\mathcal{I}}
&=&
\frac{1}{\sqrt{\Omega}}
\exp\left[
-\,i\frac{\Omega}{2}(z+z')(\tau-\tau')
\right]
\exp\left\{
-\frac{\Omega}{4}
\left[(z-z')^2+(\tau-\tau')^2\right]
\right\}
\nonumber\\
&&\times
\sum_{\ell=0}^{\infty}
\exp{\left[-S\Omega(2\ell+1)\right]}
\,L_\ell\left[
-2\,\Omega\,
(z-\mathcal{A})
(z'-\mathcal{A})
\right].
\label{Xi4}
\end{eqnarray}

It is straightforward to verify that
\begin{eqnarray}
(z-\mathcal{A})(z'-\mathcal{A})
=
-\frac{1}{4}
\left[
(z-z')^2+(\tau-\tau')^2
\right]
\equiv
-\frac{1}{4}
(\mathbf{r}_{\parallel}-\mathbf{r}_{\parallel}')^2,
\end{eqnarray}
where
\begin{eqnarray}
(\mathbf{r}_{\parallel}-\mathbf{r}_{\parallel}')^2
\equiv
(z-z')^2+(\tau-\tau')^2.
\end{eqnarray}

Therefore,
\begin{eqnarray}
\mathrm{\mathcal{I}}
&=&
\frac{1}{\sqrt{\Omega}}
\exp{\left[
i\,e\,\Phi(\mathbf{r}_{\parallel},
\mathbf{r}_{\parallel}')
\right]}
\sum_{\ell=0}^{\infty} \,\exp{\left[-\,S\,\Omega(2\ell+1)\right]}
\exp\left[
-\frac{\Omega}{4}
(\mathbf{r}_{\parallel}-\mathbf{r}_{\parallel}')^2
\right]
L_\ell
\left[
\frac{\Omega}{2}
(\mathbf{r}_{\parallel}-\mathbf{r}_{\parallel}')^2
\right].
\label{Xi5}
\end{eqnarray}
where the electric Schwinger phase is defined by
\begin{eqnarray}
\Phi
(\mathbf{r}_{\parallel},
\mathbf{r}_{\parallel}')
=-
\frac{{E}}{2}
(z+z')
(\tau-\tau').
\label{phaseE}
\end{eqnarray}

Equation~(\ref{Xi5}) may be expressed in terms of momentum variables by employing the integral representation below for the product of a Gaussian function and Laguerre polynomials~\cite{Grad}
\begin{eqnarray}
 \frac{(\alpha - \beta)^{\ell}}{\alpha^{\ell+1}}\exp\left(-{\mathcal{Z}}^{2}/2\alpha\right) L_{\ell}\left[\frac{\beta{\mathcal{Z}^2}}{2\alpha (\beta - \alpha)}\right]&=&\int_{0}^{\infty} d\mathcal{T} \,\mathcal{T} \, \exp\left(-\alpha {\mathcal{T}}^2 / 2\right) L_{\ell}\left(\beta{{\mathcal{T}}^{2}}/2\right) J_{0}(\mathcal{TZ}), \nonumber \\
\label{Gradh}
\end{eqnarray}
where $\mathcal{Z} > 0 $, $\mathrm{Re}\,(\alpha) > 0$ and $J_0$ is the Bessel function.

Also, we can rescale Eq.~(\ref{Gradh}) in an appropriate form, namely: we define 
\begin{eqnarray*}
\mathcal{T} \equiv p_{\parallel} = \sqrt{p_\tau^2 + p_z^2}, \,\,\, \mathcal{Z}^{2} \equiv ({\bf{r} - \bf{r^{\prime}}})_{\parallel}^2\,\,\,\, \mathrm{and}\,\,\,  \alpha = \beta/2 \equiv 2/\Omega.
\end{eqnarray*}
The result is 
\begin{eqnarray*}
\exp\left\{-\frac{\Omega}{4}\left[(\bf{r} - \bf{r^{\prime}})^{2}_{\parallel}\right]\right\}\, L_{\ell}\left[\frac{\Omega}{2}\,(\bf{r} - \bf{r^{\prime}})^{2}_{\parallel}\right] = \frac{2\,(-1)^{\ell}}{\Omega} \,\int_{0}^{\infty}\,dp_{\parallel}\,p_{\parallel}\,\exp\left(-\,p_{\parallel}^{2}/\Omega\right) L_{\ell}\left(2\,{p^{2}_{\parallel}}/{\Omega}\right) \,J_{0}\left(p_{\parallel}\sqrt{({\bf{r} - \bf{r^{\prime}}})_{\parallel}^{2}}\right).
\end{eqnarray*}
Substitution of this result into Eq.~(\ref{Xi5}) then gives
\begin{eqnarray}
\mathrm{\mathcal{I}} &=&\exp\left[i\,e\,\Phi(\bf{r}_{\parallel},\bf{r}_{\parallel}^{\prime})\right]\,\left(\frac{2\,\exp(-S\,\Omega)}{\sqrt{\Omega^{3}}}\right)\,\int_{0}^{\infty}\,dp_{\parallel}\,p_{\parallel}\,\exp\left(-\,p_{\parallel}^{2}/\Omega\right)\,J_{0}\left(p_{\parallel}\sqrt{({\bf{r} - \bf{r^{\prime}}})_{\parallel}^{2}}\right) \, \nonumber \\
&\times&\sum_{\ell=0}^{\infty}\,\left[-\exp\left(-\,2\,S\, \Omega\right)\right]^{\ell} \, L_{\ell}\left(2\,{p^{2}_{\parallel}}/{\Omega}\right).
\label{Xi6}
\end{eqnarray}

Now, we perform the summation over all discrete oscillator levels. The generating function for Laguerre polynomials is given by~\cite{Grad}
\begin{equation}
\sum_{\ell=0}^{\infty} \mathcal{R}^\ell L_\ell^\epsilon(Q) = \frac{1}{(1-\mathcal{R})^{1+\epsilon}} \exp\left[\frac{\mathcal{R}Q}{\mathcal{R}-1}\right],
\end{equation}
for $|\mathcal{R}| < 1$. In the present case, we have $\mathcal{R} \equiv -\exp(-2S\Omega)$ and $Q \equiv 2p_\parallel^2/\Omega$. Therefore, the summation on $\ell$ reads (for $\epsilon = 0$),
\begin{equation}
\sum_{\ell=0}^{\infty} [-\exp(-2S\Omega)]^\ell L_\ell\left(\frac{2p_\parallel^2}{\Omega}\right) = \frac{1}{1 + \exp(-2S\Omega)} \exp\left[-\frac{2p_\parallel^2}{\Omega} \frac{\exp(-2S\Omega)}{1 + \exp(-2S\Omega)}\right].
\end{equation}
By multiplying the numerator and denominator by $\exp(S\Omega)$, this expression can be rewritten as
\begin{equation}
\sum_{\ell=0}^{\infty} [-\exp(-2S\Omega)]^\ell L_\ell\left(\frac{2p_\parallel^2}{\Omega}\right) = \frac{\exp(S\Omega)}{2\cosh(S\Omega)} \exp\left[-\frac{p_\parallel^2}{\Omega} \left(1 - \tanh(S\Omega)\right)\right],
\end{equation}
which, after being substituted into Eq. (36), leads directly to
\begin{equation}
\mathcal{I} = \frac{\exp\left[ie\Phi(r_\parallel, r'_\parallel)\right]}{\sqrt{\Omega^3}\cosh(S\Omega)} \int_0^\infty dp_\parallel p_\parallel J_0\left(p_\parallel\sqrt{(\mathbf{r}-\mathbf{r}')_\parallel^2}\right) \exp\left[-\frac{p_\parallel^2}{\Omega}\tanh(S\Omega)\right].
\label{Xi7}
\end{equation}

Lastly, we are going to use the Bessel function $J_0$ in the integral representation,
\begin{eqnarray}
J_0 \left(p_{\parallel}\sqrt{({\bf{r}-\bf{r^{\prime}}})^{2}_{\parallel}}\right) = \frac{1}{2\pi}\int_{0}^{2\pi}d\varphi\exp\left[i p_{\parallel}\sqrt{({\bf{r}-\bf{r^{\prime}}})^{2}_{\parallel}} \cos\varphi\right],
\label{Bessel}
\end{eqnarray}
which allows us to write Eq.~(\ref{Xi7}) as
\begin{eqnarray}
\mathrm{\mathcal{I}} &=&2\pi\,\frac{\exp\left[i\,e\,\Phi(\bf{r}_{\parallel},\bf{r}_{\parallel}^{\prime})\right]}{\sqrt{\Omega^{3}} \,\cosh(S \, \Omega)}\,\int\,\frac{dp_{\tau}}{2\pi}\frac{dp_{z}}{2\pi}\, \exp\{i\left[p_{\tau}(\tau-\tau^{\prime})+p_{z}(z-z^{\prime})\right]\}\, \nonumber \\
&\times&\exp\left[-(p_{\tau}^{2}+p_{z}^{2})\,\frac{\tanh{(S \, \Omega)}}{\Omega}\right],
\label{Xi8}
\end{eqnarray}
where we go back to $(p_\tau,p_z)$ momenta variables, through $p_{\parallel} = (p_{\tau}^{2}+p_{z}^{2})^{1/2}$. 

Once $\mathrm{\mathcal{I}}$ has been determined, its expression can be substituted into Eq.~(\ref{16}), yielding the proper-time representation of the scalar propagator in the presence of a constant external electric field
\begin{eqnarray}
G\left(u,u^{\prime},A\right) &=& \exp\left[i\,e\,\Phi(\bf{r}_{\parallel},\bf{r}_{\parallel}^{\prime})\right]\,\int_{0}^{\infty}  dS \,\int \,\frac{d^{4}p}{(2\pi)^{4}}  \,\exp\left[ip_{\mu}\left(u-u^{\prime}\right)_{\mu}\right] \nonumber \\
&\times&\frac{1}{\cosh{(S \, \Omega)}}\,\exp\left\{-S \left[(p_{\tau}^{2}+p_{z}^{2})\frac{\tanh{(S \, \Omega)}}{S \, \Omega}+p_{x}^{2}+p_{y}^{2}+m_{0}^{2}\right]\right\}.
\label{chaves55}
\end{eqnarray}

The propagator obtained in Eq.~(\ref{chaves55}) incorporates the complete discrete spectrum generated by the external electric background. In the vanishing-field limit, $\Omega \rightarrow 0$, Eq.~(\ref{chaves55}) consistently reduces to the free scalar propagator.

It is worth noting that the scalar propagator in the presence of a constant electric field oriented along the $z$-direction is not translationally invariant in the $(\tau,z)$ sector. This behavior originates from the Schwinger phase, $\Phi(r_\parallel, r'_\parallel)$, which can be expressed through a suitable variable change involving the endpoints $u'$ and $u$. To see that, let us perform the integration of the potential connecting these endpoints, $(\tau', z')$ and $(\tau, z)$, along a straight-line path with parametrization
\begin{eqnarray*}
{A}_{\mu}({\xi}\,)  = \left(-{E}\xi_{3},0,0,0\right) \,\,\,\Rightarrow\,\,{{\bf{E}}} = -\,{\bf{\nabla}}_{{\xi}} {A}_{0}^{ext} ={{E}}\hat{z}.
\end{eqnarray*}
Following Chapter 3 of Ref.~\cite{LivroSch}, we can represent the straight line in terms of a parameter $\lambda$ as
\begin{eqnarray*}
{\mathbf{\xi}}  = (1-\lambda)\,\bf{r}_{\parallel}^{\,\prime}+\lambda\,\bf{r}_{\parallel},\,\,\,\,\,\therefore\,\,\,\,\, d\bf{\xi} = (\bf{r}_{\parallel}-\bf{r}_{\parallel}^{\,\prime})\,d\lambda, \,\,\,\,\,\,0\leq\lambda\leq1.
\end{eqnarray*}
Thus,
\begin{eqnarray}
\int_{\mathbf{r}_{\parallel}^{\prime}}^{\mathbf{r}_{\parallel}}\,{A}_{\mu}(\bf{\xi}\,) \, d{\xi_{\mu}} &=& -\int_{0}^{1}\,{E}\left[(1-\lambda)\,z^{\prime}+\lambda \,z\right](\tau-\tau^{\prime})\,d\lambda \nonumber\\
&=&-\frac{{E}}{2}(z+z^{\prime{}})(\tau-\tau^{\prime})\equiv\Phi(\mathbf{r}_{\parallel},\mathbf{r}_{\parallel}^{\prime}).
\end{eqnarray}

The phase obtained in Eq.~(44) is the endpoint-dependent Schwinger phase associated with the straight-line path connecting the two fixed points. We emphasize that this phase can be separated from the propagator without interpreting the procedure as a single global gauge transformation acting on the background field.

To isolate the Schwinger phase, we introduce the endpoint-dependent function
\begin{eqnarray}
\Lambda_{u,u'}(\xi)=\frac{{E}}{2}(z+z')\,\xi_0,
\end{eqnarray}
whose gradient integrated along the straight-line path is
\begin{eqnarray}
\int_{\mathbf{r}_{\parallel}^{\prime}}^{\mathbf{r}_{\parallel}}
\partial_{\mu}\Lambda_{u,u'}(\xi)\,
d\xi_{\mu}
=
\frac{{E}}{2}(z+z')(\tau-\tau')
=
-\Phi(\mathbf{r}_{\parallel},\mathbf{r}_{\parallel}^{\prime}).
\end{eqnarray}

Therefore,
\begin{eqnarray}
\int_{\mathbf{r}_{\parallel}^{\prime}}^{\mathbf{r}_{\parallel}}
\left[A_{\mu}(\xi)+\partial_{\mu}\Lambda_{u,u'}(\xi)\right]
d\xi_{\mu}
=
0.
\end{eqnarray}

Thus, the endpoint-dependent transformation changes the potential representation but leaves the background field strength unchanged. The vanishing of the line integral in Eq.~(47) is therefore restricted to the selected pair of endpoints and to the straight path connecting them.

Since $\Lambda_{u,u'}(\xi)$ does not modify the external electric field, the Green's function suffers the modification
\begin{eqnarray}
A_{\mu} \,\mapsto\, A_{\mu}+\partial_{\mu}\Lambda
\qquad\Longrightarrow\qquad
G\!\left(u,u',A\right)
\,\mapsto\,
G\!\left(u,u',A+\partial\Lambda\right).
\label{gaugeE}
\end{eqnarray}

Consequently, after factoring out the Schwinger phase, Eq.~(\ref{chaves55}) can be written as
\begin{eqnarray}
G\left(u-u',\Omega\right)
&=&
\int_{0}^{\infty} dS
\int\frac{d^{4}p}{(2\pi)^{4}}
\,\exp\!\left[ip_{\mu}(u-u')_{\mu}\right]
\nonumber\\
&\times&
\frac{1}{\cosh(\Omega S)}
\exp\!\left\{
-S\left[
(p_{\tau}^{2}+p_{z}^{2})
\frac{\tanh(\Omega S)}{\Omega S}
+p_{x}^{2}+p_{y}^{2}+m_{0}^{2}
\right]
\right\}.
\label{GreenTransInvE}
\end{eqnarray}

It is worth emphasizing that $G(u-u',\Omega)$ in Eq.~(\ref{GreenTransInvE}) represents the translationally invariant kernel obtained after factoring out the coordinate-dependent Schwinger phase $\Phi(r_{||}, r'_{||})$. While the full two-point propagator $G(u,u',A)$ explicitly breaks translational invariance due to the coordinate dependence of the background vector potential $A_\mu$, the isolated kernel $G(u-u',\Omega)$ preserves translational invariance in the reduced four-dimensional momentum representation.
\section{Propagation in a Uniform Electric Background: Dirac Field}
\subsection{Ritus representation of the fermionic field}

Within four-dimensional Euclidean space, the spin-$1/2$ field modes interacting with a uniform background electric configuration oriented along the $z$-axis are governed by the covariant Dirac equation
\begin{eqnarray*}
({\slashed{D}} +i\,m_{0} )\Psi(u) = 0,
\end{eqnarray*}
where we employ the minimal coupling prescription
\begin{eqnarray*}
D_{\mu} = \partial_{\mu}-ieA_{\mu}^{ext},
\end{eqnarray*}
with the external gauge potential chosen as
\begin{eqnarray*}
A_{\mu}^{ext}=(-z{E},0,0,0).
\end{eqnarray*}
The corresponding fermionic Green's function satisfies the inhomogeneous relation
\begin{eqnarray}
({\slashed{D}} +i\,m_{0})\,\mathbb{S}(u,u^{\prime},A)
=
\mathbb{I}_{4}\,
\delta^{4}(u-u^{\prime}).
\label{eqGreenElectric}
\end{eqnarray}

To establish the Clifford algebra, we adopt the Euclidean gamma matrix convention defined in Ref.~\cite{CJPEmerson2022},
\begin{equation}
\mathbb{\gamma}_{0}
=
-i
\left(
\begin{array}{cc}
\mathrm{0}_{2} & \mathbb{I}_{2}\\
\mathbb{I}_{2} & \mathrm{0}_{2}
\end{array}
\right),
\qquad
\mathbb{\gamma}_{j}
=
\left(
\begin{array}{cc}
\mathrm{0}_{2} & -\mathbb{\sigma}_{j}\\
\mathbb{\sigma}_{j} & \mathrm{0}_{2}
\end{array}
\right),
\qquad
j=1,2,3,
\label{DiracMatricesElectric}
\end{equation}
where $\mathbb{\sigma}_{j}$ denote the standard Pauli matrices. According to this representation, the components fulfill the identities
\begin{eqnarray*}
{\slashed D}
=
\gamma_{\mu}D_{\mu},
\qquad
\{\gamma_{\mu},\gamma_{\nu}\}
=
-2\delta_{\mu\nu}.
\end{eqnarray*}

Owing to the vanishing commutator between the first- and second-order Dirac operators, given by
\begin{eqnarray*}
[{\slashed D}^{2},{\slashed D}]=0,
\end{eqnarray*}
a joint basis of simultaneous eigenfunctions can be constructed. This algebraic property allows us to circumvent the direct resolution of Eq.~(\ref{eqGreenElectric}) by introducing an auxiliary Green's function $\mathbb{G}(u,u',A)$ that satisfies the second-order equation
\begin{eqnarray}
({\slashed D}^{2}+m_{0}^{2})
\,
\mathbb{G}(u,u',A)
=
\mathbb{I}_{4}
\delta^{4}(u-u').
\label{AuxiliaryElectric}
\end{eqnarray}
The fermion propagator is then recovered through
\begin{eqnarray}
\mathbb{S}(u,u',A)
=
({\slashed D}-i\,m_{0})
\,
\mathbb{G}(u,u',A).
\label{DiracFromAuxiliaryElectric}
\end{eqnarray}

Thus, it suffices to determine the auxiliary Green’s function $\mathbb{G}(u,u',A)$, from which the fermion propagator follows directly from Eq.~(\ref{DiracFromAuxiliaryElectric}).

By using the Clifford algebra properties alongside the covariant derivative commutation rule
\begin{eqnarray}
[D_{\mu},D_{\nu}]
&=&
-ieF_{\mu\nu},
\label{CommutatorElectric}
\end{eqnarray}
the quadratic Dirac operator can be expanded as
\begin{eqnarray}
\slashed D^2
&=&
-D_\mu^2
-\frac{ie}{4}
[\gamma_\mu,\gamma_\nu]F_{\mu\nu},
\end{eqnarray}
where $D_{\mu}^{2}$ represents the spinless differential operator introduced in Eq.~(\ref{operator2}).

Under the uniform electric background under consideration, the field-strength tensor possesses only the following non-vanishing elements
\begin{eqnarray}
F_{\tau z}
&=&
-\,F_{z\tau} = E.
\end{eqnarray}

Evaluating the gamma-matrix commutator explicitly within this prescribed representation yields
\begin{eqnarray*}
[\gamma_{\tau},\gamma_{z}]
=
2i \,
(\,
\tilde{\mathbb I}_{2}
\otimes
\mathbb\sigma_{z}
 \,),
\end{eqnarray*}
where the internal matrix factor is defined as
\begin{equation}
\tilde{\mathbb I}_{2}
=
\left(
\begin{array}{cc}
-1&0\\
0&1
\end{array}
\right).
\label{Itilde}
\end{equation}

This allows us to express the second-order Dirac operator in the compact form
\begin{eqnarray}
{\slashed D}^{2}
=
\Omega
\left(
\tilde{\mathbb I}_{2}
\otimes
\mathbb\sigma_{z}
\right)
-
D_{\mu}^{2}.
\label{DiracSquaredElectric}
\end{eqnarray}

Comparing the second-order Dirac operator in Eq.~(\ref{DiracSquaredElectric}) with the spinless Klein-Gordon relation in Eq.~(\ref{operator2}) demonstrates that both structures share a parallel mathematical form, differing solely by the constant matrix-valued spin-coupling term. This structural affinity suggests that the fermionic Ritus basis can be constructed systematically by incorporating a suitable spin projection operator into the scalar modes. We thus define a matrix-valued ansatz for the spinor basis elements as
\begin{eqnarray}
\mathbb{E}_{p,\sigma_{\rm E}}(u)
=
E_{p}(u)\,
\mathbf{\Xi}_{\sigma_{\rm E}},
\label{ElectricAnsatz}
\end{eqnarray}
where $E_{p}(u)$ denotes the scalar Ritus eigenfunctions given in Eq.~(\ref{RitusRepaginado}), and the projection matrix assumes the diagonal configuration
\begin{eqnarray}
\mathbf{\Xi}_{\sigma_{\rm E}}
=
{\rm diag}
\left(
\delta_{-1,\sigma_{\rm E}},
\delta_{1,\sigma_{\rm E}},
\delta_{1,\sigma_{\rm E}},
\delta_{-1,\sigma_{\rm E}}
\right),
\label{ElectricProjector}
\end{eqnarray}
with $\sigma_{\rm E}=\pm1$ labeling the eigenvalues of the matrix
$\tilde{\mathbb I}_{2}\otimes\mathbb{\sigma}_{z}$.

The projection operator satisfies the explicit eigenvalue constraint
\begin{eqnarray}
\left(
\tilde{\mathbb I}_{2}
\otimes
\mathbb{\sigma}_{z}
\right)
\mathbf{\Xi}_{\sigma_{\rm E}}
=
\sigma_{\rm E}\,
\mathbf{\Xi}_{\sigma_{\rm E}}.
\label{ElectricEigenvalue}
\end{eqnarray}
Since $\mathbf{\Xi}_{\sigma_{\rm E}}$ is a spectral projector of
$(\tilde{\mathbb I}_{2}\otimes\mathbb{\sigma}_{z})$, it commutes with this
matrix, and therefore
\begin{eqnarray}
\left(
\tilde{\mathbb I}_{2}
\otimes
\mathbb{\sigma}_{z}
\right)
\mathbb{E}_{p,\sigma_{\rm E}}(u)
=
\mathbb{E}_{p,\sigma_{\rm E}}(u)
\left(
\tilde{\mathbb I}_{2}
\otimes
\mathbb{\sigma}_{z}
\right).
\label{SpinCommutesElectric}
\end{eqnarray}

By utilizing the matrix ansatz from Eq.~(\ref{ElectricAnsatz}) alongside the projector property in Eq.~(\ref{ElectricProjector}) and the scalar Ritus basis completeness condition in Eq.~(\ref{fechamento}), coupled with the identity
\begin{eqnarray*}
\sum_{\sigma_E \,=\, \pm 1} \,\mathbf{\Xi}_{\sigma_{\rm E}} \,
\mathbf{\Xi}_{\sigma_{\rm E}}^{\dagger}
=
\mathbb I_{4},
\end{eqnarray*}
the global closure relation preserves its standard local structure:
\begin{eqnarray}
\sumint dp\,
\mathbb{E}_{p,\sigma_{\rm E}}(u)
\mathbb{E}_{p,\sigma_{\rm E}}^{\dagger}(u')
=
\mathbb I_{4}
\delta^{4}(u-u').
\label{ElectricCompleteness}
\end{eqnarray}

In analogy to the spinless sector, the existence of a complete set of eigenfunctions for the quadratic Dirac operator enables us to expand the auxiliary propagator within this basis according to
\begin{eqnarray}
\mathbb{G}(u,u^{\prime},A)
=
\sumint dp\,
\mathbb{E}_{p,\sigma_{\rm E}}(u)\,
\widetilde{\mathbb{G}}(p,A)\,
\mathbb{E}_{p,\sigma_{\rm E}}^{\dagger}(u^{\prime}).
\label{RitusExpansionElectric}
\end{eqnarray}

Following the same algebraic steps detailed for the scalar field—specifically, substituting the expansion Eq.~(\ref{RitusExpansionElectric}) into Eq.~(\ref{AuxiliaryElectric}) while enforcing the identities Eq.~(\ref{DiracSquaredElectric}), Eq.~(\ref{ElectricAnsatz}), Eq.~(\ref{SpinCommutesElectric}), and Eq.~(\ref{ElectricCompleteness})—the momentum-space auxiliary Green's function is found to inherit the matrix configuration
\begin{eqnarray}
\widetilde{\mathbb G}(p,A)
=
\frac{1}
{p^{2}
+m_{0}^{2}
+\Omega
\left(
\tilde{\mathbb I}_{2}
\otimes
\mathbb{\sigma}_{z}
\right)},
\label{MomentumPropagatorElectric}
\end{eqnarray}
where the discrete momentum squares $p^{2}$ match the spectrum given in Eq.~(\ref{pdois}).

Ultimately, inserting the momentum-space form Eq.~(\ref{MomentumPropagatorElectric}) back into the basis expansion Eq.~(\ref{RitusExpansionElectric}) completely specifies the auxiliary coordinate propagator. The physical fermion Green's function under the constant external electric background is then readily recovered by invoking Eq.~(\ref{DiracFromAuxiliaryElectric}).
\subsection{Schwinger representation of the fermionic field}

In this subsection, we focus on establishing the proper-time representation for the fermionic propagator under a constant electric background. By invoking the standard operator identity introduced in Eq.~(\ref{Operatorial}), we perform the summation over the continuous and discrete quantum states that define the electric Ritus basis. This calculation proceeds under a general algebraic structure analogous to the spinless scenario, with the unique additional modification entering via the spin interaction factor $\tilde{\mathbb I}_{2}\otimes\mathbb{\sigma}_{z}$.

As a result, the coordinate-space representation for the auxiliary Green's function is found to be
\begin{eqnarray}
\mathbb{G}\left(u,u^{\prime},A\right)
&=&
\exp\left[i\,e\,\Phi(\mathbf{r}_{\parallel},\mathbf{r}_{\parallel}^{\prime})\right]
\int_{0}^{\infty} dS
\int\frac{d^{4}p}{(2\pi)^4}
\exp\left[ip_{\mu}(u-u^{\prime})_{\mu}\right]
\nonumber\\
&\times&
\frac{1}{\cosh(\Omega S)}
\exp\left\{
-S
\left[
(p_\tau^2+p_z^2)
\frac{\tanh(\Omega S)}{\Omega S}+p_x^2+p_y^2
+m_0^2
+\Omega
(\tilde{\mathbb I}_{2}\otimes\mathbb{\sigma}_{z})
\right]
\right\}, \nonumber \\
\label{ElectricAuxiliaryProperTime}
\end{eqnarray}
where $\Omega=e E$ represents the scale established by the background field strength, and $\Phi(\mathbf{r}_{\parallel},\mathbf{r}_{\parallel}^{\prime})$ denotes the coordinate-dependent Schwinger phase characteristic of this uniform electric profile, see Eq.~(\ref{phaseE}).

By applying the first-order differential operator $(\slashed D-i\,m_0)$ directly to the auxiliary function in Eq.~(\ref{ElectricAuxiliaryProperTime}), in strict accordance with the mapping constraint Eq.~(\ref{DiracFromAuxiliaryElectric}), the intermediate coordinate expression for the full spinor propagator is found to be
\begin{eqnarray}
\mathbb{S}\left(u,u^{\prime},A\right)
&=&
\exp\left[i\,e\,\Phi(\mathbf{r}_{\parallel},\mathbf{r}_{\parallel}^{\prime})\right]
\int_{0}^{\infty} dS
\int\frac{d^{4}p}{(2\pi)^4}
\left\{
i\slashed p-im_0
\right.
\nonumber\\
&&
\left.
+i\frac{\Omega}{2}
\left[
\gamma_\tau(z-z')
-\gamma_z(\tau-\tau')
\right]
\right\}
\nonumber\\
&\times&
\frac{
\exp[ip_\mu(u-u')_\mu]
}
{\cosh(\Omega S)}
\exp
\left\{
-S
\left[(p_\tau^2+p_z^2)
\frac{\tanh(\Omega S)}{\Omega S}+p_x^2+p_y^2
+m_0^2
+\Omega
(\tilde{\mathbb I}_{2}\otimes\mathbb{\sigma}_{z})
\right]
\right\}. \nonumber \\
\label{ElectricFermionIntermediate}
\end{eqnarray}

To eliminate the explicit coordinate difference factors in favor of a momentum-space structure, we invoke a convenient differential identity acting on the internal kernel components, namely
\begin{eqnarray}
\frac{\partial}{\partial p_\tau}
\left\{
e^{ip_\tau(\tau-\tau')}
e^{-p_\tau^2\tanh(\Omega S)/\Omega}
\right\}
&=&
\left[
i(\tau-\tau')
-
\frac{2p_\tau}{\Omega}
\tanh(\Omega S)
\right]
\nonumber\\
&&\times
\left\{
e^{ip_\tau(\tau-\tau')}
e^{-p_\tau^2\tanh(\Omega S)/\Omega}
\right\},
\label{ElectricIntegralIdentityTau}
\end{eqnarray}
alongside a mathematically parallel relation for the spatial coordinate $z$. Carrying out a standard integration by parts over the momentum domains, and dropping the corresponding boundary terms due to the standard Gaussian asymptotic suppression, the expression simplifies to the more compact structure
\begin{eqnarray}
\mathbb{S}\left(u,u^{\prime},A\right)
&=&
\exp\left[i\,e\,\Phi(\mathbf{r}_{\parallel},\mathbf{r}_{\parallel}^{\prime})\right]
\int_{0}^{\infty}dS
\int\frac{d^4p}{(2\pi)^4}
\left\{
i\slashed p-im_0
\right.
\nonumber\\
&&
\left.
+
\left[
\gamma_\tau p_z
-
\gamma_z p_\tau
\right]
\tanh(\Omega S)
\right\}
\nonumber\\
&\times&
\frac{
e^{ip_\mu(u-u')_\mu}
}
{\cosh(\Omega S)}
\exp
\left\{
-S
\left[(p_\tau^2+p_z^2)
\frac{\tanh(\Omega S)}{\Omega S}+p_x^2+p_y^2
+m_0^2
+\Omega
(\tilde{\mathbb I}_{2}\otimes\mathbb{\sigma}_{z})
\right]
\right\}. \nonumber \\
\label{ElectricPropagatorIntermediate2}
\end{eqnarray}

Next, the matrix-valued spin contribution within the exponential can be explicitly unpacked through its power series expansion,
\begin{eqnarray*}
\exp
\left[-\,
\Omega S
(\tilde{\mathbb I}_{2}\otimes\mathbb{\sigma}_z)
\right]
&=&
\sum_{n=0}^{\infty}
\frac{
\left[-\,
\Omega S
(\tilde{\mathbb I}_{2}\otimes\mathbb{\sigma}_z)
\right]^n
}{n!}.
\end{eqnarray*}
By noting that the quadratic product of the spin matrices satisfies the identity $(\tilde{\mathbb I}_{2}\otimes\mathbb{\sigma}_z)^2 = \mathbb I_4$, the summation cleanly decouples into a closed hyperbolic matrix representation, yielding
\begin{eqnarray}
\exp
\left[-\,
\Omega S
(\tilde{\mathbb I}_{2}\otimes\mathbb{\sigma}_z)
\right]
&=&
\mathbb I_4\cosh(\Omega S)
-
(\tilde{\mathbb I}_{2}\otimes\mathbb{\sigma}_z)
\sinh(\Omega S).
\label{ElectricSpinIdentity}
\end{eqnarray}

Consequently, inserting Eq.~(\ref{ElectricSpinIdentity}) into the intermediate expression Eq.~(\ref{ElectricPropagatorIntermediate2}) eliminates the un-hyperbolized denominator, providing the proper-time representation for the coordinate-space fermionic Green's function:
\begin{eqnarray}
\mathbb{S}\left(u,u^{\prime},A\right)
&=&
\exp\left[i\,e\,\Phi(\mathbf{r}_{\parallel},\mathbf{r}_{\parallel}^{\prime})\right]
\int_0^\infty dS
\int\frac{d^4p}{(2\pi)^4}
\left\{
i\slashed p-im_0
\right.
\nonumber\\
&&
\left.
+
(\gamma_\tau p_z-\gamma_zp_\tau)
\tanh(\Omega S)
\right\}
\nonumber\\
&&
\times
e^{ip_\mu(u-u')_\mu}
\exp
\left\{
-S
\left[(p_\tau^2+p_z^2)
\frac{\tanh(\Omega S)}
{\Omega S}+p_x^2+p_y^2
+m_0^2
\right]
\right\}
\nonumber\\
&&
\times
\left[
\mathbb I_4
-
(\tilde{\mathbb I}_{2}\otimes\mathbb{\sigma}_z)
\tanh(\Omega S)
\right].
\label{ElectricProperTimePropagator}
\end{eqnarray}
The relation derived in Eq.~(\ref{ElectricProperTimePropagator}) defines the exact charged fermion propagator under a constant external Euclidean electric background. As expected, the coordinate dependence is isolated within the Schwinger phase factor $\Phi(\mathbf{r}_{\parallel},\mathbf{r}_{\parallel}^{\prime})$, which explicitly couples the endpoints of the parallel sectors, while the remaining operational kernel depends strictly on coordinate differences.

By isolating the translational non-invariant component through exactly the same prescription employed in the previous section for the scalar field case, which is constructed from the initial and final parallel space-time points along a straight line, the spinor propagator can be written as 
\begin{eqnarray}
\mathbb{S}(u-u',\Omega)
&=&
\int_0^\infty dS
\int\frac{d^4p}{(2\pi)^4}
\left\{
i\slashed p-im_0
+
(\gamma_\tau p_z-\gamma_zp_\tau)
\tanh(\Omega S)
\right\}
\nonumber\\
&&\times
e^{ip_\mu(u-u')_\mu}
\exp
\left\{
-S
\left[(p_\tau^2+p_z^2)
\frac{\tanh(\Omega S)}
{\Omega S}+p_x^2+p_y^2
+m_0^2
\right]
\right\}
\nonumber\\
&&\times
\left[
\mathbb I_4
-
(\tilde{\mathbb I}_{2}\otimes\mathbb{\sigma}_z)
\tanh(\Omega S)
\right].
\label{ElectricGaugeIndependentPropagator}
\end{eqnarray}
Similarly to the scalar case, the fermionic kernel $\mathbb{S}(u-u',\Omega)$ in Eq. (\ref{ElectricGaugeIndependentPropagator}) constitutes the translationally invariant part of the Dirac propagator.


\section{Comparison with the Literature}

In the formal vanishing-field limit where $ E \rightarrow 0$ (or equivalently $\Omega \rightarrow 0$), the substitution $\tanh(\Omega S) \rightarrow \Omega S$ takes place, and Eq.~(\ref{ElectricGaugeIndependentPropagator}) successfully reduces back to the free Euclidean fermion propagator, as physically required.

It is also useful to comment on the connection between the formulation using the Euclidean electric field ${E}$ and the Minkowski electric field $E_{\rm{M}}$. The present expressions are obtained in terms of the Euclidean field. Upon the appropriate analytic continuation to the Minkowski electric field, the hyperbolic functions appearing in Eqs.~(\ref{GreenTransInvE}) and (\ref{ElectricGaugeIndependentPropagator}) are transformed into the trigonometric structures characteristic of the Schwinger propagator. In particular, the substitution ${E}\rightarrow -\,iE_{\rm{M}}$ implies $\Omega\rightarrow -i\,eE_{\rm{M}}$, so that
\begin{equation}
\tanh(\Omega S)\rightarrow -\,i\tan(eE_{\rm{M}}S),
\label{74}
\end{equation}
up to the convention adopted for the Wick rotation. This provides a direct connection between the representation using the Euclidean field ${E}$ and the description of charged fields in a constant Minkowski electric background $E_{\rm{M}}$ as adopted in Refs.~\cite{Xu,Xu2,EeB}.

On the other hand, we can use the propagator obtained here to compute the effective Lagrangian of the system. To see that, consider the expression given by Eq.~($1$) from~\cite{ElmforsPRL}, namely
\begin{eqnarray}
\frac{\partial \mathcal{L}_{\mathrm{eff}}}{\partial m_0} = i\lim_{u\rightarrow u^{\prime}} \,\mathrm{Tr}\left[\mathbb{S}\left(u,u^{\prime},A^{ext}\right)\right],
\label{1}
\end{eqnarray}
where $\mathbb{S}\left(u,u^{\prime},A^{ext}\right)$ is the Dirac propagator computed in Eq.~(\ref{ElectricProperTimePropagator}) and $\mathrm{Tr}$ is the trace operation over fermion indices. Integrating Eq.~(\ref{1}), we obtain
\begin{eqnarray}
\mathcal{L}_{\mathrm{eff}} = i\lim_{u\rightarrow u^{\prime}}\int\,\mathrm{Tr}\left[\mathbb{S}\left(u,u^{\prime},A^{ext}\right)\right]\,dm_0 + const.
\label{2}
\end{eqnarray}
The integration constant does not depend on the fermion mass and, for simplicity, is set to zero. The trace operation over  $\mathbb{S}\left(u,u^{\prime},A^{ext}\right)$ is performed in Appendix A [see Eq.~(\ref{eq:final_trace_result}). Thus, we obtain
\begin{eqnarray*}
\mathcal{L}_{\mathrm{eff}}
&=& -\,2 \, \int_0^\infty \frac{dS}{S}
\int\frac{d^4p}{(2\pi)^4} \,
\exp
\left\{
-S
\left[(p_\tau^2+p_z^2)
\frac{\tanh(\Omega S)}
{\Omega S}+p_x^2+p_y^2
+m_0^2
\right]
\right\}.
\label{ElectricGaugeIndependentPropagator-P}
\end{eqnarray*}

Performing the Gaussian integrals, we get
\begin{eqnarray*}
\mathcal{L}_{\mathrm{eff}}
&=&  -\,\frac{1}{8\pi^2}\, \int_0^\infty \frac{dS}{S^2} \,\exp
\left(
-S \,m_0^2
\right)\,\frac{\Omega}{\tanh{(\Omega \,S)}}.
\label{ElectricGaugeIndependentPropagator-P}
\end{eqnarray*}

Let us renormilize {\textit{à lá Schwinger}}. The final expression for effective Lagrangian becomes
\begin{eqnarray}
\mathcal{L}_{\mathrm{eff}}^{ren}
&=&  -\,\frac{1}{8\pi^2}\, \int_0^\infty \frac{dS}{S^3} \,\exp
\left(
-S \,m_0^2
\right)\,\left\{\Omega \,S\,\coth{(\Omega \,S)}-1-\frac{(\Omega S)^2}{3}\right\}.
\label{ElectricGaugeIndependentPropagator-P}
\end{eqnarray}

To transition to the physical Minkowski spacetime, we implement the analytical continuation prescription given by Eq.~(\ref{74}) into Eq.~(\ref{ElectricGaugeIndependentPropagator-P}), resulting in
\begin{eqnarray}
\mathcal{L}_{\mathrm{eff}}^{\rm{ren}}
&=&  -\,\frac{1}{8\pi^2}\, \int_0^\infty \frac{dS}{S^3} \,\exp
\left(
-S \,m_0^2
\right)\,\left\{(eE_{M} S)\,\cot{(e E_{M} S)}-1+\frac{(e E_{M} S)^2}{3}\right\}.
\label{ElectricGaugeIndependentPropagator-PPP}
\end{eqnarray}

Note that $\cot(e E_M S)$ has poles at:
\begin{equation}
    S_n = \frac{n \pi}{e E_M}.
    \label{poles}
\end{equation}

We can expand $f(S)$ around the poles $S_n$:
\begin{eqnarray*}
    f(S) = f(S_n) + \left. \frac{df(S)}{dS} \right|_{S=S_n} (S - S_n) + \cdots,
\end{eqnarray*}
for both $f(S) = \cos(e E_M S)$, and independently of $f(S) = \sin{e E_M S}$. Then, we obtain
\begin{equation}
    \cot(e E_M S) \approx \sum_{n=1}^{\infty} \frac{1}{e E_M (S - S_n)} + \mathcal{O}\left( S - S_n \right).
\end{equation}

Now, by shifting the poles $S_n \to S_n - i\epsilon$ and using the Sokhotski--Plemelj formula:
\begin{align}
    \lim_{\epsilon \to 0^+} \frac{1}{e E_M (S - S_n + i\epsilon)} &= \lim_{\epsilon \to 0^+} \frac{1}{e E_M (S - S_n) + i\epsilon} \nonumber \\
    &= \mathcal{P}\left( \frac{1}{e E_M (S - S_n)} \right) - i \pi \delta\Big( e E_M (S - S_n) \Big),
\end{align}
reads
\begin{eqnarray}
    \cot(e E_M S) \approx \sum_{n=1}^{\infty} \left[ \mathcal{P}\left( \frac{1}{e E_M (S - S_n)} \right) - \frac{i\pi}{e E_M} \delta(S - S_n) \right],
    \label{cot}
\end{eqnarray}
where $\mathcal{P}$ denotes the Cauchy principal value.

Replacing Eq.~(\ref{cot}) on Eq.~(\ref{ElectricGaugeIndependentPropagator-PPP}), we have
\begin{align}
\mathcal{L}_{\mathrm{eff}}^{\rm{ren}}
&=  -\,\frac{1}{8\pi^2}\, \int_0^\infty \frac{dS}{S^3} \,\exp
\left(
-S \,m_0^2
\right)\,\left\{(eE_{M} S) \frac{}{}\right.\nonumber \\ 
&\times\left.\left[\sum_{n=1}^{\infty} \left[ \mathcal{P}\left( \frac{1}{e E_M (S - S_n)} \right) - \frac{i\pi}{e E_M} \delta(S - S_n) \right]\right]-1+\frac{(e E_{M} S)^2}{3}\right\}.
\label{ElectricGaugeIndependentPropagator-PPPP}
\end{align}

Since we are interested in the rate of pair production per unit of spacetime, which is defined as $\Gamma = 2\,\mathrm{Im}(\mathcal{L}_{\mathrm{eff}}^{\mathrm{ren}})$, we get the following real-valued expression
\begin{align*}
\mathrm{Im}(\mathcal{L}_{\mathrm{eff}}^{\mathrm{ren}}) = \frac{\pi}{8\pi^2}\, \sum_{n=1}^{\infty}\,\frac{\exp{(-S_n m_{0}^{2})}}{S_{n}^2}.
\end{align*}
Substituting the poles (\ref{poles}) into the last equation, we finally find the pair production rate per unit of spacetime
\begin{align}
\Gamma_{\rm{pairs}} = \frac{1}{4\pi} \,\sum_{n=1}^{\infty}\,\left(\frac{eE_M}{n \pi}\right)^{2}\,\exp{\left(-\frac{n \pi m_{0}^{2}}{eE_M}\right)}.
\label{rate}
\end{align}

Equation \eqref{rate} reproduces the well-known nonperturbative
result for pair production in spinor QED under a constant electric
field, originally derived by Schwinger. It agrees term by term
with Eq.~(6.41) of Ref.~\cite{Schwinger}, providing a consistency
check of the hybrid Ritus--Schwinger scheme developed in this work.

\section{Comments and Conclusions}
In this paper, we have presented a description of both spinless and spinor quantum field propagators under a uniform, constant external electric background. By combining the spectral diagonalization properties of the Ritus eigenfunction technique with the operator language of Schwinger’s proper-time framework, we have derived exact analytical representations for charged bosonic and fermionic Green’s functions in coordinate space. This integrated scheme combines the advantages of both approaches, successfully carrying out the discrete sum over the complete set of eigenfunctions while yielding a compact proper-time kernel.

A central advantage of our derivation is that the exact Green's functions retain their four-dimensional momentum representation. This feature ensures that all momentum degrees of freedom are explicitly preserved for subsequent applications, thereby providing an unconstrained field-theoretic framework to introduce thermal and finite-volume modifications through generalized momentum compactification. The translationally invariant components isolated after extracting the coordinate-dependent Schwinger phase serve as the essential starting blocks needed to formulate quantum field dynamics within toroidal topologies in the presence of constant electric backgrounds.

Comparing the findings of this paper with the standard Schwinger results presented in Ref.~\cite{Schwinger}, we emphasize the following key points:
 \begin{itemize}
        \item Schwinger's original operator representation expresses the phase coordinate dependence through line integrals $\exp\left[ie \int_{x''}^{x'} dx_\mu A_\mu(x)\right]$. In our hybrid scheme, this non-translationally invariant phase is explicitly factorized as $\Phi(r_\parallel, r_\parallel') = -\frac{E}{2}(z+z')(\tau-\tau')$.
        \item While Schwinger's expression encodes field interactions via matrix tensor contractions $\exp\left[\frac{i}{4}(x'-x'')\,eF\coth(eFS)(x'-x'')\right]$, our Ritus spectral summation yields an explicit 4D momentum distribution. The transverse momentum components $p_\perp = (p_x, p_y)$ remain uncoupled with standard Gaussian weight $\exp[-S(p_x^2+p_y^2+m_0^2)]$, whereas the longitudinal components $p_\parallel = (p_\tau, p_z)$ parallel to the electric field are modulated by $\frac{\tanh(\Omega S)}{\Omega S}$.
        \item The internal matrix-valued spin structure in our Euclidean formulation, $\exp\left[-\,
\Omega S
(\tilde{\mathbb I}_{2}\otimes\mathbb{\sigma}_z)
\right]$, directly maps onto Schwinger's electromagnetic spin-coupling tensor $\exp\left[i\frac{e}{2}\left(\sigma \cdot F\right) S\right]$.
    \end{itemize}

The hybrid approach implemented here does not use the full operator manipulations often associated with Schwinger’s conventional proper-time derivation while simultaneously circumventing the specialized coordinate restrictions originally required by the Ritus method for the Dirac field. Under the hybrid approach considered in this contribution, the limit $\Omega \rightarrow 0$ correctly yields the free propagators for both spin-0 and spin-1/2 fields; furthermore, we successfully rederive the pair production rate for the fermionic field.

\section*{Acknowledgments} 
I would like to thank Leon Sandbote for the discussions.
\subsection{Trace Evaluation of Propagator}
\label{subsec:trace_evaluation}

To analyze specific radiative corrections or effective actions, it is often necessary to evaluate the Dirac trace of the proper-time propagator kernel. Let us denote the matrix structure inside the integrand of Eq.~\eqref{ElectricProperTimePropagator} as $\mathbf{M}(p, S)$. Utilizing the Euclidean conventions established in Eq.~\eqref{DiracMatricesElectric}, this kernel is explicitly written as:
\begin{equation}
\mathbf{M}(p, S) = \left\{ i\slashed p - im_0 + \Gamma_{\tau z} \tanh(\Omega S) \right\} \left[ \mathbb I_4 - \Sigma_z \tanh(\Omega S) \right],
\label{eq:matrix_M}
\end{equation}
where we have defined the shorthand operators for the longitudinal momentum structures and the spin projection as:
\begin{equation}
\Gamma_{\tau z} \equiv \gamma_\tau p_z - \gamma_z p_\tau, \qquad \Sigma_z \equiv \tilde{\mathbb I}_{2} \otimes \mathbb{\sigma}_z.
\end{equation}

Before computing the Dirac trace, we apply the distributive property to expand the product within the brackets, yielding six distinct algebraic terms:
\begin{align}
\mathbf{M}(p, S) &= i\slashed p - i\slashed p \Sigma_z \tanh(\Omega S) - im_0 \mathbb I_4 + im_0 \Sigma_z \tanh(\Omega S) + \Gamma_{\tau z} \tanh(\Omega S) - \Gamma_{\tau z} \Sigma_z \tanh^2(\Omega S).
\label{eq:distributive_expansion}
\end{align}

Taking the trace ($\mathrm{Tr}$) over the $4 \times 4$ spinorial space, the linearity of the trace operation allows us to evaluate each contribution independently:
\begin{itemize}
    \item $\mathrm{Tr}(i\slashed p) = 0$ and $\mathrm{Tr}[\Gamma_{\tau z} \tanh(\Omega S)] = 0$, since the trace of any linear combination of single Euclidean gamma matrices vanishes identically.
    
    \item $\mathrm{Tr}[im_0 \Sigma_z \tanh(\Omega S)] = 0$, because the Pauli matrix $\mathbb{\sigma}_z$ is traceless ($\mathrm{Tr}\,\mathbb{\sigma}_z = 0$), which implies $\mathrm{Tr}\,\Sigma_z = \mathrm{Tr}(\tilde{\mathbb I}_{2}) \times \mathrm{Tr}(\mathbb{\sigma}_z) = 2 \times 0 = 0$.
    
    \item $\mathrm{Tr}[i\slashed p \Sigma_z \tanh(\Omega S)] = 0$. In terms of $2 \times 2$ blocks, the momentum operator contains only off-diagonal elements, $\slashed p = \left(\begin{smallmatrix} \mathbf{0}_2 & A \\ B & \mathbf{0}_2 \end{smallmatrix}\right)$, with $A = ip_\tau \mathbb{I}_2 - \vec{p}\cdot\vec{\sigma}$ and $B = ip_\tau \mathbb{I}_2 + \vec{p}\cdot\vec{\sigma}$. Computing the direct product with the block-diagonal spin operator in the `align` environment yields:
    \begin{align}
    i\slashed{p}\Sigma_z &= i \begin{pmatrix} \mathbf{0}_2 & A \\ B & \mathbf{0}_2 \end{pmatrix} \begin{pmatrix} \sigma_z & \mathbf{0}_2 \\ \mathbf{0}_2 & \sigma_z \end{pmatrix} 
    = \begin{pmatrix} \mathbf{0}_2 & i A \sigma_z \\ i B \sigma_z & \mathbf{0}_2 \end{pmatrix}.
    \label{eq:align_p_slash}
    \end{align}
    Since the main diagonal consists strictly of $\mathbf{0}_2$ blocks, the trace vanishes identically.
    
    \item $\mathrm{Tr}[\Gamma_{\tau z} \Sigma_z \tanh^2(\Omega S)] = 0$. Similarly, the longitudinal field operator maps into anti-diagonal blocks, $\Gamma_{\tau z} = \left(\begin{smallmatrix} \mathbf{0}_2 & C \\ D & \mathbf{0}_2 \end{smallmatrix}\right)$, where $C = -ip_z \mathbb{I}_2 + p_\tau \sigma_z$ and $D = -ip_z \mathbb{I}_2 - p_\tau \sigma_z$. Evaluating the product via blocks reveals:
    \begin{align}
    \Gamma_{\tau z}\Sigma_z &= \begin{pmatrix} \mathbf{0}_2 & C \\ D & \mathbf{0}_2 \end{pmatrix} \begin{pmatrix} \sigma_z & \mathbf{0}_2 \\ \mathbf{0}_2 & \sigma_z \end{pmatrix} = \begin{pmatrix} \mathbf{0}_2 & C \sigma_z \\ D \sigma_z & \mathbf{0}_2 \end{pmatrix}.
    \label{eq:align_gamma_long}
    \end{align}
    As the resulting matrix possesses exclusively off-diagonal block structures, the main $4 \times 4$ diagonal contains no non-zero elements, confirming that $\mathrm{Tr}[\Gamma_{\tau z}\Sigma_z] = 0$.
\end{itemize}

Unifying these results, the only non-vanishing contribution arises purely from the mass term proportional to the identity matrix $\mathbb I_4$. The exact algebraic evaluation of the proper-time kernel trace reduces to:
\begin{equation}
\mathrm{Tr}\left[ \mathbf{M}(p, S) \right] = \mathrm{Tr}\left( -i \,m_0 \,\mathbb I_4 \right) = -\,4\,i\,m_0.
\label{eq:final_trace_result}
\end{equation}
%

\end{document}